\documentstyle[prl,preprint,aps,epsf]{revtex}

\def\addcontentsline#1#2#3{\relax}

\begin{document}

\title{\Large\bf ``Flux'' state in double exchange model}
\author{Masanori Yamanaka, Wataru Koshibae$^1$, and Sadamichi Maekawa$^1$}
\address{Department of Applied Physics, Science University of Tokyo, 
Kagurazaka, Shinjuku-ku, Tokyo 162, Japan}
\address{${}^1$Institute for Material Research,  
Tohoku University, Katahira, Sendai, 980-8577, Japan}
\date{April 28, 1998}
\maketitle

\begin{abstract}
We study the ground state properties of the double-exchange systems.
The phase factor of the hopping amplitude arises 
from $t_{2g}$ spin texture in two or more dimensions.
A novel ``flux'' state has a lower energy than the canted 
antiferromagnetic and spiral spin states. 
In a certain range of hole doping, 
a phase separation occurs between the ``flux'' state
and antiferromagnetic states.
Constructing a trial state 
which provides the rigorous upper bound on the ground state,
we show that the metallic canted antiferromagnetic state
is not stable in the double exchange model.
\end{abstract}

\newpage

The double-exchange model 
\cite{REFzener,REFanderson,REFdegennes}
has received special attention since the discovery 
of the colossal magnetoresistance (CMR) in manganites
\cite{REFexperiments}.
The {\em least common} model includes a single tight-binding band
of $e_g$ electrons with the nearest neighbour hopping ($t$)
coupled with the localized $t_{2g}$ spins ($S$) 
by the strong Hund coupling ($J_H$).
The $t_{2g}$ spins interact each other 
by the antiferromagnetic (AF) exchange interaction ($J$).
The hopping amplitude depends on the relative directions 
of the localized magnetic moments.
Anderson and Hasegawa \cite{REFanderson} derived the elements 
from the analysis of two sites problem.
Intuitively, they are the inner product of the classical spin vectors 
on the adjacent sites. 

De Gennes obtained \cite{REFdegennes} the phase diagram 
as a function of temperature ($T$) and hole concentration ($x$). 
At zero temperature and $x$$=$$0$, the system exhibits the AF phase.
The ferromagnetic phase away from $x$$=$$0$ was attributed 
to the double-exchange mechanism.
The existence of the canted AF phase between them 
is an interesting and remarkable phenomenon induced by 
a competition between the kinetic energy of $e_g$ electrons 
and the AF exchange coupling among $t_{2g}$ spins 
\cite{REFnagaev}.
The canted angle changes continuously between the two phases
as a function of doping.
These properties have been theoretically conceptualized
as fundamental and inherent properties of the double-exchange model.
Recently, Yunoki et al. obtained the phase diagram 
by an intensive numerical study.
The results suggest the existence of a phase separation
\cite{REFyunokimoreo}.
(See also \cite{REFkkm}.)
In the other publications, 
it was claimed that one needs an additional phase factor 
in the hopping amplitude \cite{REFdm},
and several properties associated to it have been reported
\cite{REFcb,REFkht,REFkmms}.

In this letter,
we investigate the ground state properties 
of the double-exchange model in two dimension
in the large limits of $J_H$ and $S$.
The model is reduced to a tight-binding model with complex hopping
and with a single band.
The phase factor of hopping arises from the $t_{2g}$ spin texture 
in two or more dimensions \cite{REFdm}.
(In one dimension, it can be always eliminated by a local gauge 
transformation and was missed in two sites problem \cite{REFanderson}.)
We focus on the phase degrees of freedom, 
which makes remarkable effects,
and obtain the following results:
(i) A novel ``flux'' state has a lower energy than 
the canted AF \cite{REFdegennes}
and spiral spin states \cite{REFinouemaekawa,comment:qvector}. 
This state has a commensurate spin texture.
We have optimized {\em the shape of the density of states}
in addition to the band width and the exchange energy
because the flux may induce an extension of the Brillouin zone
or the Fermi surface instability (gap opening). 
The mechanism is similar to the flux phase discussed 
in the Hubbard model \cite{REFam,REFlieb} 
and the generalized Peierls instability \cite{REFhlrw}
exhibited by lattice fermions in a magnetic field.
(ii) In a certain range of $x$, 
the phase separation occurs between 
the ``flux'' and AF states.
We propose a fundamental mechanism of the phase separation.

Due to the above effects, it is concluded that 
the metallic canted AF state
is unstable in the double-exchange model.
More precisely, we explicitly construct a trial state
whose energy is lower than that of the canted AF state.
It provides the {\em rigorous upper bound} on the ground state energy,
which is sufficient to prove the absence of the canted AF state
for a wide parameter region.
The dimensionality (two or more) plays an essential role 
and the mechanism is completely distinct from
that observed in one dimension \cite{REFyunokimoreo,REFonedim}.
These results are stable under the perturbation 
with respect to the three dimensional layered structures
if the inter-layer hoppings are small enough.
The recent neutron scattering experiments\cite{REFky,REFhirotaendoh}
and the transport properties 
are discussed in the light of the present results.

The double-exchange model at finite doping
is defined by
\begin{eqnarray} 
H=-t \sum_{\langle i, j \rangle} 
\Big[
\Big(
\cos{\frac{\theta_i}{2}}
\cos{\frac{\theta_j}{2}}
+ e^{-i ( \phi_i - \phi_j)}
\sin{\frac{\theta_i}{2}}
\sin{\frac{\theta_j}{2}}
\Big) 
c^{\dagger}_i c^{\phantom{\dagger}}_j + h.c.
\Big]
+J 
\sum_{\langle i, j \rangle} 
\vec{S}_i \cdot \vec{S}_j,
\label{eq:hamiltonianKJ}
\end{eqnarray} 
where $c_i$ is the annihilation operator of spinless fermion,
the summation is over nearest-neighbor sites in the square lattice,
and $i$ (or $j$) denotes the lattice site.
The second term is the AF exchange coupling between $t_{2g}$ localized 
spins, where $\vec{S}_i$ is the classical spin directing
$(\theta_i, \phi_i)$ in the polar coordinate \cite{comment:canted}.
The shape of the density of states depends on the phase factor 
in the hopping.
Therefore, to study the ground state, we have to optimize the band width, 
the exchange energy, and the shape of the density of states
\cite{comment:degennes1}.
It should be noted that any analysis which assume an ordering pattern 
of two sublattices misses the possibility of the ``flux'' state 
{\em a priori}, because the effect of the phase factor originates 
from the global structure of $t_{2g}$ spins. 
We begin with parametrization of the phase factor
in the hopping amplitude assuming the 
$\sqrt{2}$$\times$$\sqrt{2}$ structure of the $t_{2g}$ spin texture.
The extension to bigger structures is straightforward.
We choose the parametrization of the $t_{2g}$ spin texture
\begin{eqnarray}
(\theta_i, \phi_i) =
\left\{
\begin{array}{@{\,}lll}
(\theta_1, \phi)        & \mbox{for} &  i=(2m,2n)     \\
(\theta_2, \phi+\frac{\pi}{2})  & \mbox{for} & i=(2m+1,2n)   \\
(\theta_1, \phi+\pi)    & \mbox{for} & i=(2m+1,2n+1) \\
(\theta_2, \phi+\frac{3\pi}{2}) & \mbox{for} & i=(2m,2n+1)  
\end{array}\right.,
\label{eq:parameter}
\end{eqnarray}
where $m$ and $n$ are integers.
The ferromagnetic and AF states are expressed by
$(\theta_1, \theta_2, \phi)$$=$$(0,0,\phi)$ and 
$(\theta_1, \theta_2, \phi)$$=$$(0,\pi,\phi)$, respectively.
The $e_g$ electrons acquire a flux 
whose magnitude is equivalent to the surface of the unit sphere
surrounding by the four vectors (\ref{eq:parameter}) 
in a motion around the plaquette.
The Brillouin zone is extended to two sites per unit cell
due to the phase degrees of freedom.
The dispersion relation is 
\begin{eqnarray}
E_{\pm}=\pm 2t 
\sqrt{
A^2 (\cos{k_x}+\cos{k_y})^2
+
B^2 (\cos{k_x}-\cos{k_y})^2
},
\nonumber
\end{eqnarray}
where $A$$=$$\cos{\theta_1}\cos{\theta_2}$ 
and $B$$=$$\sin{\theta_1}\sin{\theta_2}$.
The Brillouin zone is 
$(k_x, k_y)$ $\in$ $\{ |k_x +k_y | \le \pi \}$ 
$\cap$ $\{ |k_x - k_y | \le \pi \}$.
The corresponding density of states is 
\begin{eqnarray}
\rho(E)=
\left\{
\begin{array}{@{\,}l}
\frac{8}{\pi^2}
\frac{1}{| E |}
K\left(
1-\frac{4(\alpha E^2 -\gamma)}{E^4}
\right)^{1/2}
\nonumber\\
\hspace{24mm}
\mbox{for} \ 
\gamma/\alpha
< E^2 < 2(\alpha-\beta)\\
\frac{2}{\pi^2}
\sqrt{
\frac{E^2}{\alpha E^2 - \gamma}
}
K\left(1-
\frac{E^4}{4(\alpha E^2 -\gamma)}
\right)^{1/2}
\nonumber\\
\hspace{36mm}
\mbox{for} \ 
2(\alpha-\beta) < E^2 \\
\frac{8}{\pi^2}
\sqrt{
\frac{E^2}{4 (\gamma - \alpha E^2 ) +E^4}
}
K\left(1-
\frac{4\alpha \gamma}
{4\alpha \gamma +E^4}
\right)^{1/2}
\nonumber\\
\hspace{43mm}
\mbox{for} \
E^2 < 
\gamma/\alpha
\end{array}\right.,
\nonumber
\end{eqnarray}
where $K(x)$ is the complete elliptic integral,
$\alpha$$=4t^2(A^2+B^2)$, 
$\beta$$=4t^2(A^2-B^2)$, 
and $\gamma$$=64t^4A^2B^2$. 
(See Fig.~\ref{FIGUREdos}.)
The splitting of the density of states corresponds
to an extension of the Brilliouin zone.
The formation of the (pseudo-)energy gap is crucial to stabilize 
the ``flux'' state against the canted AF state.  
The state gains the total energy optimizing the commensurate spin 
structure through the phase degrees of freedom
(a Peierls phase).

To obtain the phase diagram, we compare the total energy 
among the states exhibited 
by the Hamiltonian (\ref{eq:hamiltonianKJ}) with 
the parametrization (\ref{eq:parameter}).
The total energy is optimized in the parameter space,
$JS^2$, $\theta$'s, and $\phi$, setting $t$$=$$1$.
(It should be noted that the energy is invariant under 
the global $U(1)$ transformation with respect to $\phi$,
which plays an important role in the stability 
for the three dimensional layered structures as we will show below.)
The canted AF state is not included in the (\ref{eq:parameter})
and we obtain the energy from the model (\ref{eq:hamiltonianKJ})
with $\phi_i$$=$$0$ and $\theta_i$$=$$\theta$
where the canted angle is specified by the $\theta$.
The results are shown in Fig.~\ref{FIGUREdiagram}(a).
(I) {\em near half filling}:
For $JS^2$$\stackrel{\large{>}}{\sim}$$0.13$,
the ``flux'' state has a lower energy than the ferromagnetic,
canted AF, and spiral spin states.
In Fig.~\ref{FIGUREenergy}(a), we show energy of the states 
as a function of $x$ at $JS^2$$=$$0.3$. 
As $JS^2$ decreases, a transition to the ferromagnetic state occurs.
This transition is due to a level crossing between them. 
In Fig.~\ref{FIGUREenergy}(b), we show energy of the states 
as a function of $x$ at $JS^2$$=$$0.1$. 
The configuration of spin angles 
is shown in Fig.~\ref{FIGUREspinconfig}.
As $JS^2$ increases, the configuration asymptotically
approaches to that of the AF state.
(II) {\em low doping region}:
A ``flux'' state with 
$(\theta_1, \theta_2, \phi)$$=$$(0, \theta, \phi)$ and 
the canted AF state \cite{REFdegennes} degenerate, 
but these states are unstable against the phase separation 
as we show below.

The density of states of the ferromagnetic, canted AF, and 
spiral states has one van Hove singularity at the band center,
while the ``flux'' state has two singularities and a pseudo-gap.
When the Fermi level is in the pseudo-gap
the ``flux'' state gains its energy.
We have to note the difference of unit cell
between the hopping amplitude of $e_g$ electron
and the ordering structure of the $t_{2g}$ spins.
The number of sites in their unit cells 
{\em are not} necessarily the same.
For example, the unit cell of the $t_{2g}$ spins has
two sites for the canted AF state, and a lot of sites 
for the spiral spin states.
On the other hand, the unit cell of $e_g$ electrons
has only one site for the canted AF, spiral, ferromagnetic
spin states.

So far, we restricted the spin structure to be 
$\sqrt{2}$$\times$$\sqrt{2}$.
The optimized state has a staggered ``flux'' structure
because the total phase around a plaquette by (2)
is opposite in sign for two neighbor plaquettes.
The state is, so called, the staggered flux state
and is distinct from the flux state in the usual sense.
Therefore, we use the double quotation to distinguish it.
Generalizing the mechanism 
of the $\sqrt{2}$$\times$$\sqrt{2}$ ``flux'' state, 
a flux state characterized by a bigger unit cell structure
could have a lower energy than the canted AF state
at an arbitrary rational filling. 
Because, if we assume a unit cell structure $a$$\times$$b$$\equiv$$q$
($a$, $b$$>$$\sqrt{2}$),
the structure allows to suppress the density of states 
at $x$$=$$p$$/$$q$ where $p$ and $q$ are coprimers.
Especially, when we restrict the discussion 
inside the ``Flux'' region in Fig.~\ref{FIGUREdiagram}(a),
the $a$$\times$$b$ flux state could gain some energy 
with respect to the $\sqrt{2}$$\times$$\sqrt{2}$ ``flux'' state
except at half-filling. 
Although we have not explored in such bigger structures in detail, 
it would be interesting to study the relation among
the flux (or ``flux'') state, the rational hole concentration,
and a bigger structure of the unit cell 
along the line of the present paper.

Assuming the ordering of the $t_{2g}$ spins, the phase diagram 
is obtained (Fig.~\ref{FIGUREdiagram}(a)).
However, the canted AF phase is unstable under phase separation.
The phase diagram is shown in Fig.~\ref{FIGUREdiagram}(b).
As shown in Figs.~\ref{FIGUREenergy},
the system can gain the energy 
by mixing the AF and the state whose energy level touches the 
tangent line from the AF point, i.e. $x$$=$$0$ \cite{comment:degennes2}.
For $JS^2 \stackrel{\large{>}}{\sim} 0.23$,
it exhibits the phase separation with the AF and ``flux'' spin states.
For $JS^2 \stackrel{\large{<}}{\sim} 0.17$,
the system exhibits the phase separation 
between the AF and ferromagnetic states.
A fraction of the pure canted AF state survives
in Fig.~\ref{FIGUREdiagram}(b)(iii),
but the state degenerates to a ``flux'' state 
with $(\theta_1, \theta_2, \phi)$$=$$(0, \theta, \phi)$ there.
The nature of the boundaries between the phase separation 
and the ordered state is expected to be of the first order.
The existence of the ``flux'' state is crucial to
the absence of the canted AF state.
Otherwise the canted AF state still remains instead of 
the region of the ``flux'' state in Fig.~\ref{FIGUREdiagram}(a),
and the phase separation between AF and canted AF occurs
in the region $JS^2 \stackrel{\large{>}}{\sim} 0.23$.
Although the $\sqrt{2}$$\times$$\sqrt{2}$ ``flux'' state is a 
trial state in the sense of the variational argument,
it provides the {\em rigorous upper bound} on the ground state energy.
Therefore, this is sufficient to prove the absence
of the canted AF state in the wide parameter region.
(It is possible that the true ground state is a ``flux'' state 
with bigger structures or another state
and the phase separation between the state and AF state occurs.)

The ``flux'' state is stable against the perturbations
with respect to three dimensional layered structures.
This is due to the following reasons:
(i) the structure of the van Hove singularity in each layer is stable
if the inter-layer hoppings are small enough.
(ii) Each layer has the global $U(1)$ symmetry with respect to $\phi$, 
and can adjust the AF exchange energy taking different $\phi$'s. 
The quantitative study is a problem in future.

Experimentally, the ``flux'' state and the phase separation 
between the state and AF state are particularly interesting.
The spin configuration of the ``flux'' state is unusual 
\cite{comment:qvector}.
In the neutron scattering experiments,
this will be detected as three spots,
$(\pi/2, \pi/2)$, $(0, \pi)$, and $(\pi, 0)$ 
corresponding a complex folding of the Brillouin zone.
A remnant of the phase separation between the ``flux'' 
and AF states may be detected 
by a crossover of the intensities as a function of $x$,
because the intensities of each component 
can be evaluated independently.
However, one cannot distinguish the phase separation 
from the other phenomena only from the neutron scattering
(see \cite{REFky,REFhirotaendoh,REFallodi}).
The $a$$\times$$b$ flux state could gain some energy 
with respect to the $\sqrt{2}$$\times$$\sqrt{2}$ ``flux'' state
in the ``Flux'' region in Fig.~\ref{FIGUREdiagram}(a)
or to the canted AF in the other region.
It is expected that the stabilization occurs with an energy gap
due to a similar mechanism to \cite{REFhlrw}
or due to the mechanism \cite{REFonedim}.
These conditions meet the quantization of the Hall conductance
\cite{REFtknn,REFcomment}.

In summary, we have investigated the universal properties 
of the double-exchange mechanism at zero temperature.
The ``flux'' state has a lower energy than the canted AF
and spiral states.
The phase degrees of freedom and the dimensionality 
(two or more) play an essential role.
It is concluded that the metallic canted AF state
is unstable in the double exchange model
within the analysis of the {\it least common} model, i.e. 
without any assistance of other effects like polaron
\cite{REFkkm}, orbitals, etc. 
This is one of few examples of the ``flux'' state \cite{REFfradkin}
realized in non-singular model.

We thank Masaki Oshikawa for his important contribution 
to one-dimensional models
and for kindly allowing us to mention the resluts in this letter.
We are grateful to Yasuo Endoh, Kazuma Hirota, Sumio Ishihara, 
Satoshi Okamoto, and Yoshinori Tokura for useful discussions.
This work was supported by CREST and a Grant-in-Aid 
from the Ministry of Educations, Science and Culture in Japan.
The computation in this work has been done using the facilities
of the Supercomputer Center, Institute for Solid State Physics,
University of Tokyo 
and Institute for Material Science, Tohoku University.

\begin{figure}
\caption{
Densities of states in two dimension.
The solid line is for the ``flux'' state.
The broken line is for the ferromagnetic, canted AF,
and spiral spin states.
}
\label{FIGUREdos}
\end{figure}

\begin{figure}
\caption{
(a) Phase diagram of the double exchange model, which exhibits
the ``flux'' (``Flux'') and ferromagnetic (F).
The canted AF state and ``flux'' state with
$(\theta_1, \theta_2, \phi)$$=$$(0, \theta, \phi)$ 
degenerate (CAF). 
(b) Phase diagram after taking into account the phase separation,
which exhibits the phase separation 
between AF and ``flux'' (AF+``Flux''), 
AF and ferromagnetic (AF+F), 
AF and the canted AF (AF+CAF),
``flux'' and Ferromagnetic (i),
``flux'' and canted AF (ii) spin states.
In (iii) the canted AF and ``flux'' states degenerate.
}
\label{FIGUREdiagram}
\end{figure}

\begin{figure}
\caption{
Energy of the ``flux'' state (i), canted AF state (ii),
ferromagnetic states (iii) 
as a function of $x$ at $JS^2$$=$$0.3$ (a) and  $JS^2$$=$$0.1$ (b).
The tangent line (iv) from $x$$=$$0$ is an energy of phase separation.
}
\label{FIGUREenergy}
\end{figure}

\begin{figure}
\caption{Energy dependence of $\theta's$
as a function of $J$ at $x$$=$$0.4$.}
\label{FIGUREspinconfig}
\end{figure}


\begin{thebibliography}{99}
%
\bibitem{REFzener}
C.~Zener,  
Phys. Rev. {\bf 82}, 403 (1951). 

\bibitem{REFanderson}
P.W.~Anderson and H.~Hasegawa:
Phys. Rev. {\bf 100}, 675 (1955).

\bibitem{REFdegennes}
P.-G.~de Gennes, Phys. Rev. {\bf 118}, 141 (1960).

\bibitem{REFexperiments}
For the perovskite structure such as 
A$_{1-x}$B$_x$MnO$_3$, see, for example,  
S.~Jin, T.H.~Tiefel, M.~McCormack, R.A.~Fastnacht, 
R.~Ramesh, and L.H.~Chen, Science {\bf 264}, 413 (1994).
For the layered compounds described by
(A,B)$_{n+1}$Mn$_n$O$_{3n+1}$, see, for example, 
R.A.~Moham Ram, P.~Ganguly, and C.N.R.~Rao,
J. Solid State Chem. {\bf 70}, 82 (1987).

\bibitem{REFnagaev}
For the canted AF state, see also
\'E.L.~Nagaev, Soviet Phys. JETP, {\bf 30}, 693 (1970).

\bibitem{REFyunokimoreo}
S.~Yunoki, H.~Hu, A.~Malvezzi, A.~Moreo, N.~Furukawa, and E.~Dagotto,
Phys. Rev. Lett.{\bf 80}, 845 (1998);
S.~Yunoki, A.~Malvezzi, A.~Moreo, H.~Hu, S.~Capponi, D.~Poilblanc
and N.~Furukawa, cond-mat/9709029;
Yunoki and Moreo, cond-mat/9712152.

\bibitem{REFkkm}
M.Yu.~Kagan, D.I.~Khomskii, and M.~Mostovoy,
cond-mat/9804213.

\bibitem{REFdm}
E.~M\"uller-Hartman and E.~Dagotto, 
Phys. Rev. B{\bf 54}, R6819 (1996).

\bibitem{REFcb}
M.J.~Calder\'on and L.~Brey,
cond-mat/9801311.

\bibitem{REFkht}
H.~Koizumi, T.~Hotta, and Y.~Takada,
cond-mat/9801093.

\bibitem{REFkmms}
Y.B.~Kim, P.~Majumdar, A.J.~Millis, and B.I.~Shraiman,
cond-mat/9803350.

\bibitem{REFinouemaekawa}
J.~Inoue and S.~Maekawa, Phys. Rev. Lett. {\bf 74}, 3407 (1995).

\bibitem{comment:qvector}
The spin configuration of the ``flux'' state cannot 
be specified by a single $Q$ vector. 
In this sense, the state is distinct from the spiral spin state.

\bibitem{REFam}
I.~Affleck and J.B.~Marston, Phys. Rev. B{\bf 37}, 3774 (1988).

\bibitem{REFlieb}
E.H.~Lieb, Phys. Rev. Lett. {\bf 73}, 2158 (1994).

\bibitem{REFhlrw}
Y.~Hasegawa, P.~Lederer, T.M.~Rice, and P.B.~Wiegmann,
Phys. Rev. Lett. {\bf 63}, 907 (1989).

\bibitem{REFonedim}
In one dimension, there is essentially no phase degree of freedom.
However, the hopping amplitude in the Hamiltonian 
(\ref{eq:hamiltonianKJ}) can induce Peierls instability
(formation of energy gaps) by using $\theta$'s degrees of freedom,
i.e. $t_{2g}$ spin textures.
This is expected to be essential properties in one dimension. 
For example, the alternation of the hopping means 
that there are {\em two sites} in unit cell for $e_g$ electron,
which corresponds to {\em four-fold} periodicity 
in unit cell for $t_{2g}$ spins. 
(See the text for the difference of the unit cells.)
The enhancement of spin structure factor at $k$$=$$\pi/2$ 
in \cite{REFyunokimoreo} is due to this property.
These arguments are based on private communications 
with Masaki Oshikawa and the resluts including higher dimensions 
will be published elsewhere. 

\bibitem{REFky}
H.~Kawano. R.~Kajimoto, M.~Kubota, and H.~Yoshizawa,
Phys. Rev. B{\bf 53}, 14709 (1996). 

\bibitem{REFhirotaendoh}
K.~Hirota, N.~Kaneko, A.~Nishizawa, Y.~Endoh,
M.C.~Martin, and G.~Shirane, Physica B{\bf 237-238}, 36 (1997). 

\bibitem{comment:canted}
Usually, the $\phi$'s are ignored
and the hopping amplitude is reduced
to $\cos(\theta_i-\theta_j)/2$.
On the basis, the canted AF and the spiral spin configurations
give the same hopping amplitude
and we never distinguish each other.

\bibitem{comment:degennes1}
In \cite{REFdegennes,REFkkm}, the hopping amplitude is real 
and the optimization is needed only for the band width 
(of the {\em single} cosine band) and the exchange energy.

\bibitem{comment:degennes2}
In \cite{REFdegennes,REFkkm}, 
the phase diagram was obtained by a perturbation 
with respect to $x$, i.e. in the low doping limit.

\bibitem{REFallodi}
G.~Allodi, R.~De Renzi, G.~Guidi, F.~Licci and, M.W.~Pieper,
Phys. Rev. B{\bf 56}, 6036 (1997).

\bibitem{REFtknn}
D.J.~Thouless, M.~Kohmoto, M.P.~Nightingale, and M.~den Nijs,
Phys. Rev. Lett. {\bf 49}, 405 (1982).

\bibitem{REFcomment}
The Hamiltonian (\ref{eq:hamiltonianKJ}) is obtained 
by a unitary transformation.
It is interesting how the quantization of the Hall conductance 
in (\ref{eq:hamiltonianKJ}) does relate to the Hall effects
in the original model. 

\bibitem{REFfradkin}
For example, see E.~Fradkin, 
{\em Field Theories and Condensed Matter Systems}, 
Addison-Wesley (1991).

\end{thebibliography}
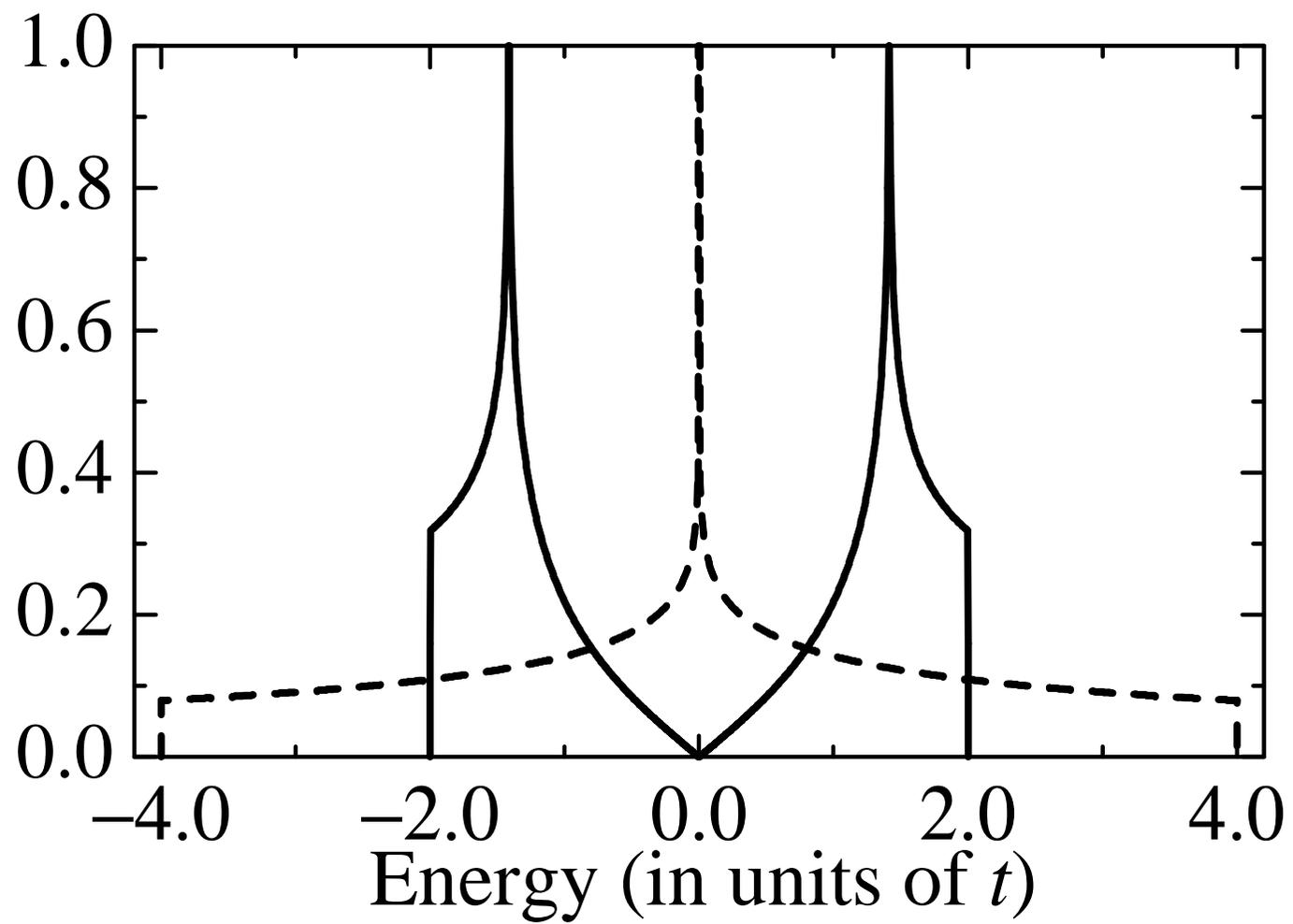
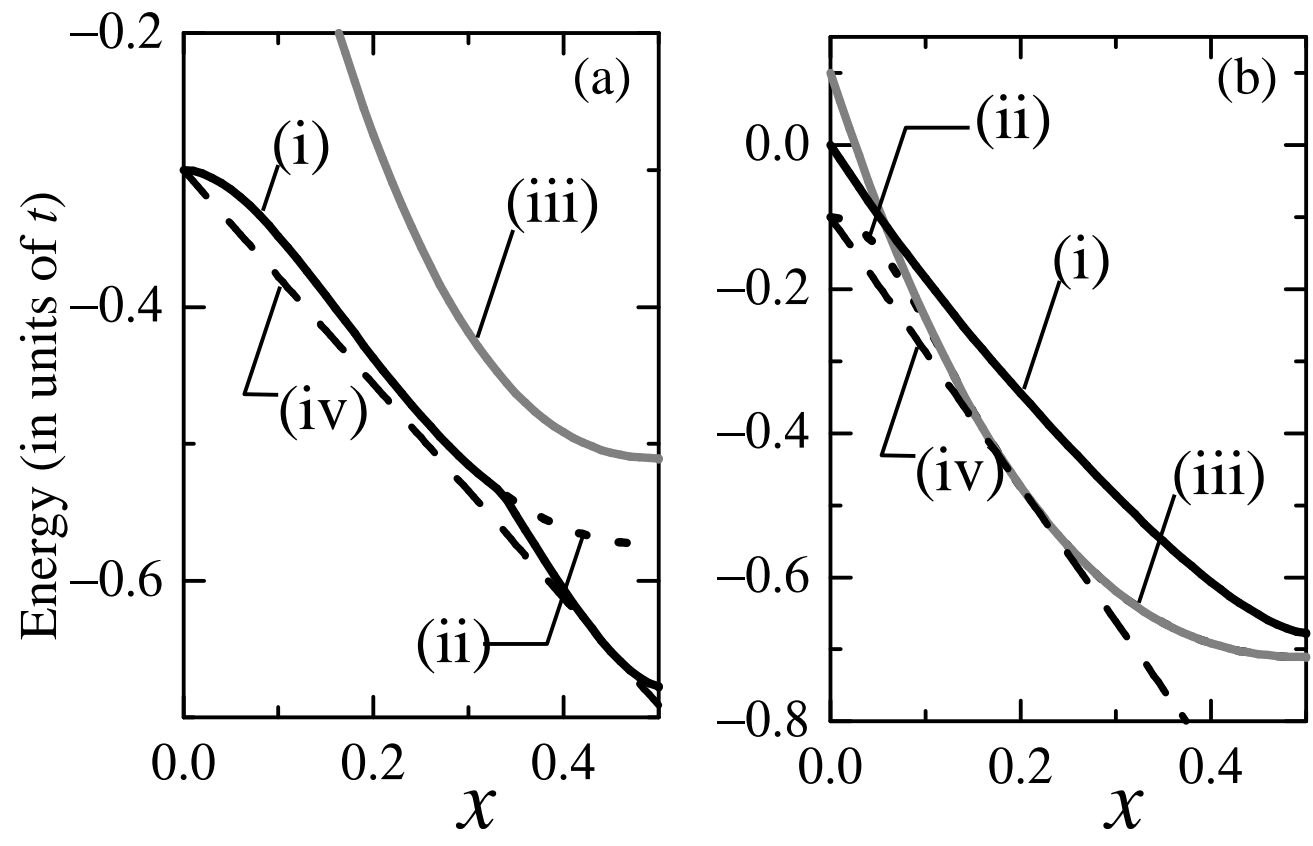
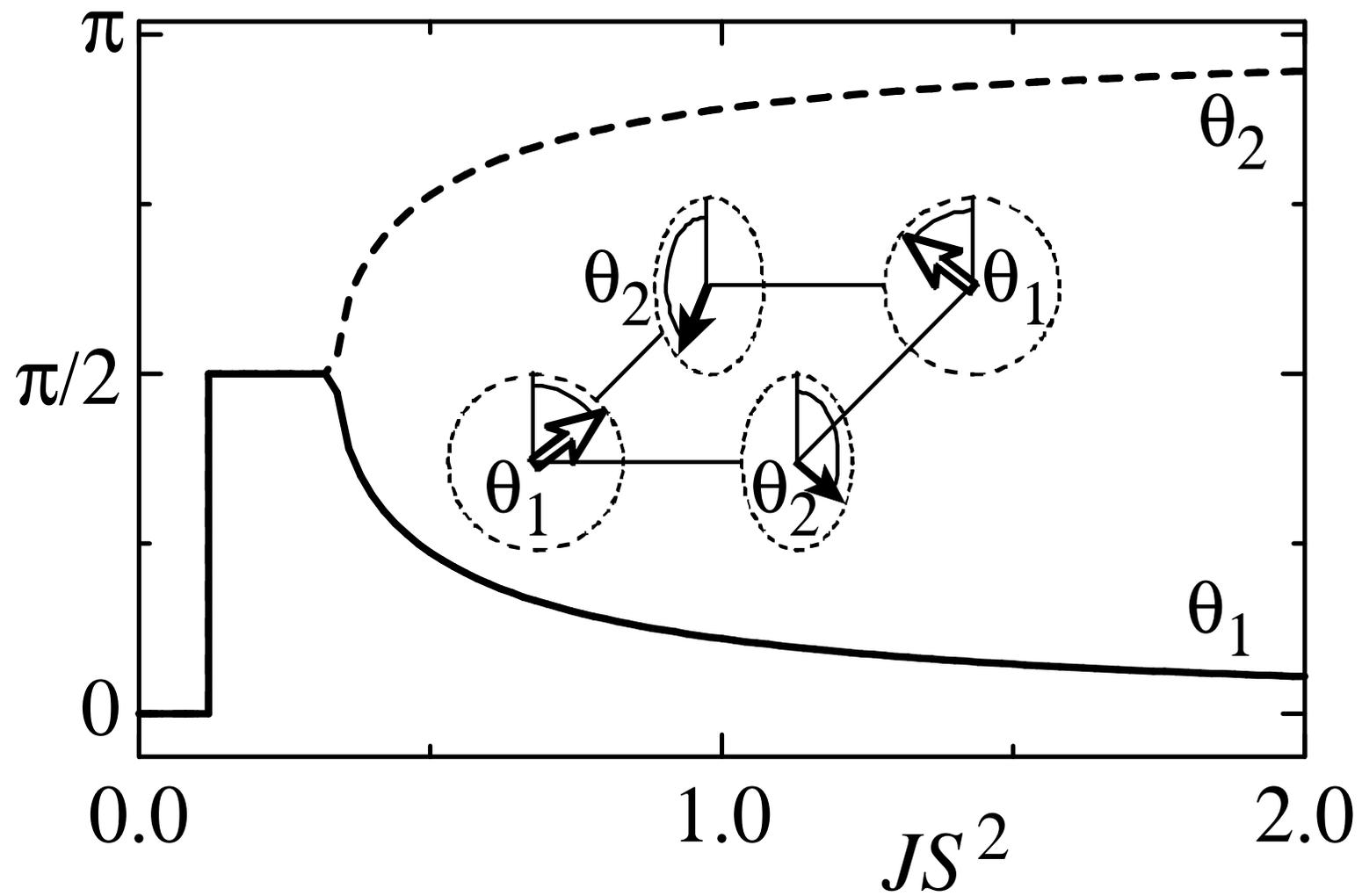
\end{document}